\documentclass[letter]{aa}     % for the letters
\usepackage{graphicx}
\usepackage{txfonts}
\usepackage{lipsum}
\usepackage{subcaption}         % necessary for continued figures, example in section 3
\usepackage{lscape}             % to rotate a single page table, example in appendix.
\usepackage{placeins}           % useful with \FloatBarrier, to keep 
\newcommand{\chandra}{{\it Chandra}}

\newcommand{\xmm}{{\it XMM}-Newton}

\newcommand{\swift}{{\it Swift}}

\newcommand{\tess}{{\it TESS}}

\newcommand{\yzret}{YZ~Ret}

\begin{document}

%%%%%%%%%%%%%%%%%%%%%%%%%%%%%%%%%%%%%%%%
% if you use custom commands in your title,
% ensure to check your title when submitting!
%%%%%%%%%%%%%%%%%%%%%%%%%%%%%%%%%%%%%%%%
   %\title{Detection of a 37.7 s modulation in the optical light curve of Nova Ret 2020. (This title will change)}
%   \title{The nature of the fast optical modulation in Nova Ret 2020: a stable 37.7 s white dwarf spin.}
\title{A very rapidly rotating white dwarf in nova YZ Reticuli}
   %\subtitle{Subtitle}

%%%%%%%%%%%%%%%%%%%%%%%%%%%%%%%%%%%%%%%%
% Please separate each author with the \and command
%
% Use the \corrauth to provide the corresponding
% author address. It will be automatically inserted as 
% footnote in the PDF output.
%
% Please DO NOT include ORCIDs next to author names.
% Instead, please provide an active address for each coauthor:
% it will be automatically extracted by EDPS editorial system, 
% and co-authors will be be able to authenticate their ORCID.
%
% Only authenticated ORCIDs will be taken into account.
% ORCIDs included here will be removed.
%%%%%%%%%%%%%%%%%%%%%%%%%%%%%%%%%%%%%%%%

   \author{G. J. M. Luna\inst{1,2}\corrauth{juan.luna@unahur.edu.ar}        % use \corrauth for the corresponding author
                \and N. Rawat\inst{3}
                \and R. Angeloni\inst{4}
                \and M. Orio\inst{5,6} 
                \and S. Scaringi\inst{7,8}
                \and A. Dobrotka\inst{9} 
                \and J. Magdolen\inst{9}
                %\and D. Buckley\inst{...}
                %\and S. Potter\inst{...}
        }

   \institute{Universidad Nacional de Hurlingham (UNAHUR). Laboratorio de Investigación y Desarrollo Experimental en Computación, Grupo de Astrofísica, Av. Gdor. Vergara 2222, Villa Tesei, Buenos Aires, Argentina
   \and Consejo Nacional de Investigaciones Científicas y Técnicas (CONICET), Argentina.
   \and South African Astronomical Observatory, PO Box 9, Observatory, Cape Town, 7935, South Africa
   \and International Gemini Observatory/NSF NOIRLab, Casilla 603, La Serena, Chile
  \and INAF-Osservatorio Astronomico di Padova, vicolo Osservatorio, 5, 35122 Padova, Italy
  \and Department of Astronomy, University of Wisconsin, 475 N. Charter Str., Madison WI 53706, WI, USA
    \and Centre for Extragalactic Astronomy, Department of Physics, Durham University, South Road, Durham, DH1 3LE, UK
    \and INAF -- Osservatorio Astronomico di Capodimonte, Salita Moiariello 16, I-80131 Naples, Italy
    \and Advanced Technologies Research Institute, Faculty of Materials Science and Technology in Trnava, Slovak University of Technology in Bratislava, Bottova 25, 917 24 Trnava, Slovakia    
}

   \date{Received ..., 20XX}

  \abstract
{YZ~Ret (Nova Reticuli 2020) is the first VY~Scl-type nova-like variable observed to undergo a classical nova eruption. Following the outburst, timing analysis of 20-s cadence \tess\ data revealed a periodicity at approximately 42\,s, suggesting a possible classification as a fast-spinning Intermediate Polar. To definitively identify the nature of this modulation, we performed a multi-instrument timing analysis using high-speed ground-based photometry Zorro/Gemini South (1\,s cadence) and the South African Astronomical Observatory (5\,s cadence) alongside \tess\ Sector 97 observations. Our ground-based data reveal a highly coherent period of 37.69131$\pm$0.00001\,s, which we identify as the true rotation period of the white dwarf. We demonstrate that the apparent 42.61\,s signal in the \tess\ data is a Nyquist alias of this fundamental frequency. Furthermore, the signal amplitude in the \tess\ data is suppressed by a factor of $\approx$0.6 relative to the Gemini observations, a result consistent with the theoretical sinc-function damping expected for a 20-s integration time. The extreme coherence and long-term stability of the 37.69131\,s signal rule out transient phenomena such as dwarf nova oscillations or non-radial pulsations. We conclude that \yzret\ hosts a fast-spinning magnetic white dwarf in an Intermediate Polar configuration. This discovery implies that mass loss during the nova eruption was likely driven by a fast magnetic rotator wind and provides a physical explanation for the missing supersoft X-ray phase, suggesting that nearly the entire accreted envelope was exhausted, promptly quenching the nuclear burning.}


   \maketitle
\nolinenumbers

%%%%%%%%%%%%%%%%%%%%%%%%%%%%%%%%%%%%%%%%%%%%%%%%%%%%%%%%%%%%%%
\section{Introduction} \label{sec:intro}

\yzret\ 
(MGAB-207) showed  the ``first'' of several astrophysical phenomena: before the outburst, it was one of the few novae already known as cataclysmic variables (CVs); it was classified as a VY~Scl-type nova-like after its brightness was reported to fade for about 2 magnitudes for prolonged periods of time \citep[][]{Murawski2019}; early in July 2020, \yzret\ was discovered as a bright, "naked-eye" classical nova, when it reached V$\approx$3.7 \citep[see][and references therein]{Sokolovsky_YZRET}; and thanks to the low column density towards the source, it was the nova where the long-predicted initial X-ray flash of the ``fireball'' phase was first detected \citep{Konig2022}.

Early post-outburst observations revealed a system in rapid transition, characterized by intense super-soft X-ray emission and complex optical variability. It is important to note that the super-soft luminous X-ray source was never observed, and we will return to this point in the section where we discuss our conclusions. This was tentatively explained with a high inclination, so that the central source was obscured by the disk \citep{Sokolovsky_YZRET, Mitrani2024} and permitted to analyze the shocks occurring at a late phase of the outburst in the nova ejecta, which collided with a thin shell of previously ejected cold and dense material, producing a spectrum with several features from radiation recombination continua and the best evidence -- and the first case in a nova -- of an astrophysical object outside the Sun undergoing the charge exchange phenomenon \citep[see][]{Mitrani2024}. 

In this Letter, we present compelling evidence of the magnetic nature of the white dwarf through a comprehensive timing analysis. We utilize a combination of space-based photometry from the Transiting Exoplanet Survey Satellite (\tess) and ground-based Gemini South (GS) and $SAAO$ high-speed photometry to resolve the nature of the system's periodicity in the high frequency range. In Section \ref{sec:obs}, we describe the data and analysis techniques. Section \ref{sec:res} presents our results, while Section \ref{sec:disc} interprets them in the context of the nova aftermath. In Section \ref{sec:concl} we summarize our conclusions.

\begin{figure*}
    \centering
    \includegraphics[scale=0.45]{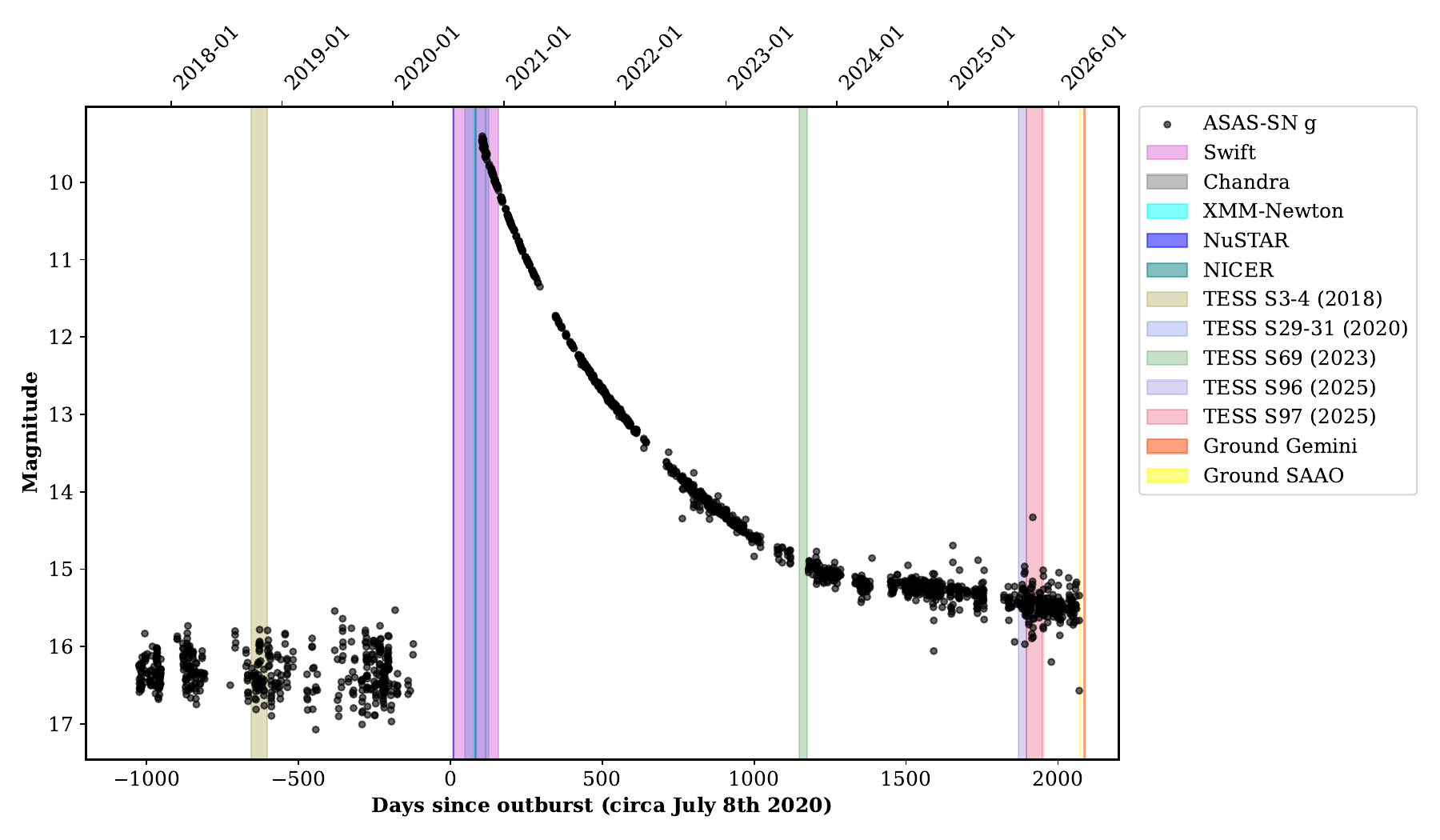}
    \caption{Multi-wavelength long-term light curve of \yzret\ covering the period 2017-2026, starting in quiescence until the most recent observations. The panel shows the optical evolution from ASAS-SN. Vertical colored regions indicate the epochs of high-cadence \tess\ observations analyzed by \citet{Schaefer2022} and in this work (\tess\ Sectors 3, 4, 29, 30, 31, 69, 96, and 97), X-ray observations from \swift, \chandra, and \xmm, and high-speed ground-based photometry from GS/$Zorro$ and $SAAO$. The light curve illustrates the transition from the post-outburst decline to the current VY~Scl-type high state, providing the temporal context for the detected white dwarf spin.}
    \label{fig:context}
\end{figure*}

\section{Observations and data analysis} \label{sec:obs}

\subsection{\tess}

\yzret\ was observed by \tess\ in Sectors 3, 4, 29, 30, 31 and 69; these data were already analyzed and reported by \citet{Schaefer2022}. It was observed in Sectors 96 and 97 more recently (PI: Scaringi). In this article, we concentrate on the observation of sector 97, which has a 20\,s cadence light curve available (PI: Scaringi) in the Barbara A. Mikulski Archive for Space Telescopes (MAST)\footnote{\url{https://archive.stsci.edu }}. The observations of Sector 97 started on September 15 2025 and lasted for 54.62 days. \yzret\ at this time was still about 1-$g$ magnitude above the pre-nova quiescent level, as shown in Fig. \ref{fig:context}, where we plot the $g$-magnitude light curve from the ASAS-SN archive \citep{Kochanek2017}. 

\subsection{South African Astronomical Observatory/SHOC}
Observations of \yzret\ were obtained in the {\em V} band with the 1.9-m telescope equipped with the Sutherland High-Speed Optical Camera (SHOC) at the South African Astronomical Observatory (SAAO), Sutherland. Individual exposures of 5\,s were acquired using GPS-triggered timing. The target was monitored on 11 and 13 March 2026, with an average coverage of approximately 1.1\,h per night. Data reduction followed standard procedures implemented in \texttt{Python}, primarily using \texttt{photutils} \citep[v2.2.0;][]{Photutils2016} together with the Astropy-affiliated package \texttt{astroquery.gaia} \citep{astropy:2013,astropy:2018}. Instrumental magnitudes were extracted via aperture photometry. Photometric calibration was performed using two field stars with magnitudes in the {\em Gaia} archive\footnote{\url{https://gea.esac.esa.int/archive/}}. Synthetic $g$-band magnitudes derived from {\em Gaia} photometry were used to determine the photometric zero-point and calibrate the target light curve.

\subsection{Gemini South/Zorro}
\yzret\ was observed with the Zorro fast dual-channel camera\footnote{https://www.gemini.edu/instrumentation/alopeke-zorro} mounted on the Gemini South telescope (Scott et al. 2021; Howell et al. 2025). The observations were carried out on 26 March 2026 as part of the Director’s Discretionary Time program GS-2026A-DD-105 (PI: G. J. M. Luna). Zorro was operated in wide-field mode, with the target positioned at the center of the $\sim$35 arcsec diameter unvignetted field of view. The blue and red channels were configured with $2\times2$ binning and a $1024\times1024$ pixel window, using the XSDSS $g^\prime$ filter, centered at 479 nm, and the XSDSS $i^\prime$ filter, centered at 765 nm, respectively. The EMCCD detectors were operated as conventional CCD imagers, acquiring a two-hour time series consisting of four sequences of 1750 frames each, for a total of 7000 frames per channel, with a cadence of 1\,s. The observations were obtained under photometric conditions, with an average seeing of $\sim$0.8$^{\prime\prime}$.

Because no suitable comparison star was available within the Zorro wide-field of view, aperture photometry was performed on the target alone after standard bias and flat-field corrections. The target counts were extracted within a circular aperture of radius 2$^{\prime\prime}$ and analyzed using \texttt{AstroImageJ} (Collins et al. 2017). The $g^\prime$ and $i^\prime$ light curves were analyzed independently and produced consistent results; for clarity, only the $g^\prime$ light curve is shown in Figure \ref{fig:mirror}.

\subsection{Timing analysis}
We constructed the periodograms of each light curve using the Lomb-Scargle (LS) algorithm as implemented in the \texttt{astropy} library \citep{Astropy2022}, and adopted its standard normalization. We focused on the frequencies above 0.005\,Hz, those not sampled in the previous analysis of \tess\ data by \citet{Schaefer2022}. In this frequency region, the LS periodogram was dominated by white noise, thus we applied the standard determination of the False Alarm Probability at the 99.9\% level. The uncertainty of the detected period was estimated using a Monte Carlo approach combined with a residual bootstrapping technique. After identifying the strongest peak suspected to be the primary frequency in the original Lomb-Scargle periodogram, we subtracted the corresponding best-fit sinusoidal model from the data to isolate the residuals. We then generated \(N=1,000\) synthetic light curves by adding the best-fit model to a randomized version of the residuals, obtained through bootstrapping. The periodogram was re-calculated for each synthetic dataset, and the formal uncertainty was defined as the standard deviation of the resulting distribution of periods. 

\section{Results} \label{sec:res}

Our analysis of the \tess\ Sector 97 periodogram in the frequency range above 0.005\, Hz and up to $f_{\text{Nyquist}}$=0.025\,Hz revealed a strong peak in the frequency $f_{TESS}$=0.023468\,Hz ($P_{TESS}$=42.61050 $\pm$ 0.00003\,s). Given the 20\,s cadence of the \tess\ data and the "uncomfortably" close proximity of $f_{TESS}$ to $f_{\text{Nyquist}}$, before going further into its interpretation in the context of \yzret, we obtained the observations with $SAAO$ and GS described in the previous sections. With 5\,s and 1\,s cadences, we extended the Nyquist limit to 0.1 and 0.5\,Hz, respectively. 

\begin{figure}
    \centering
    \includegraphics[scale=0.35]{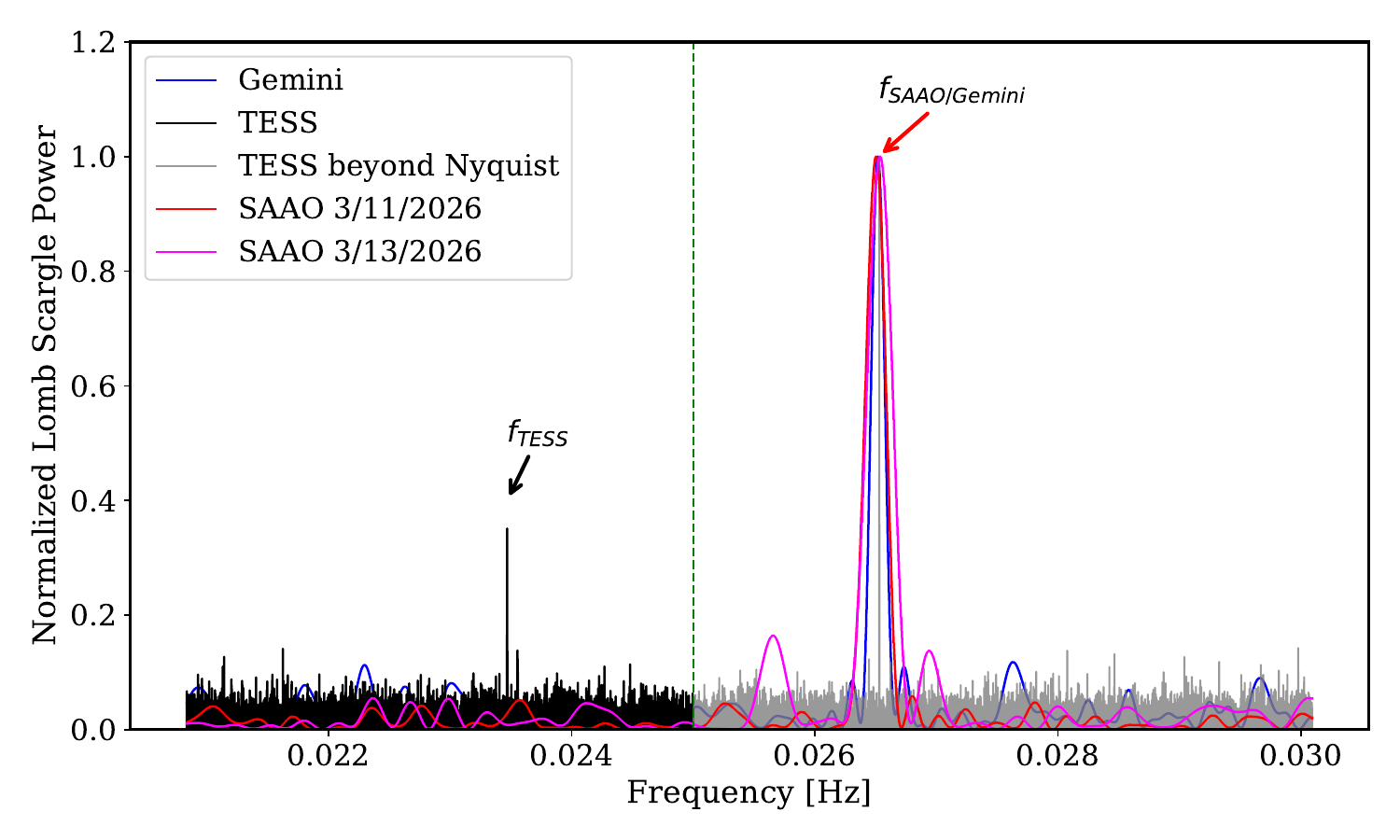}
    \includegraphics[scale=0.35]{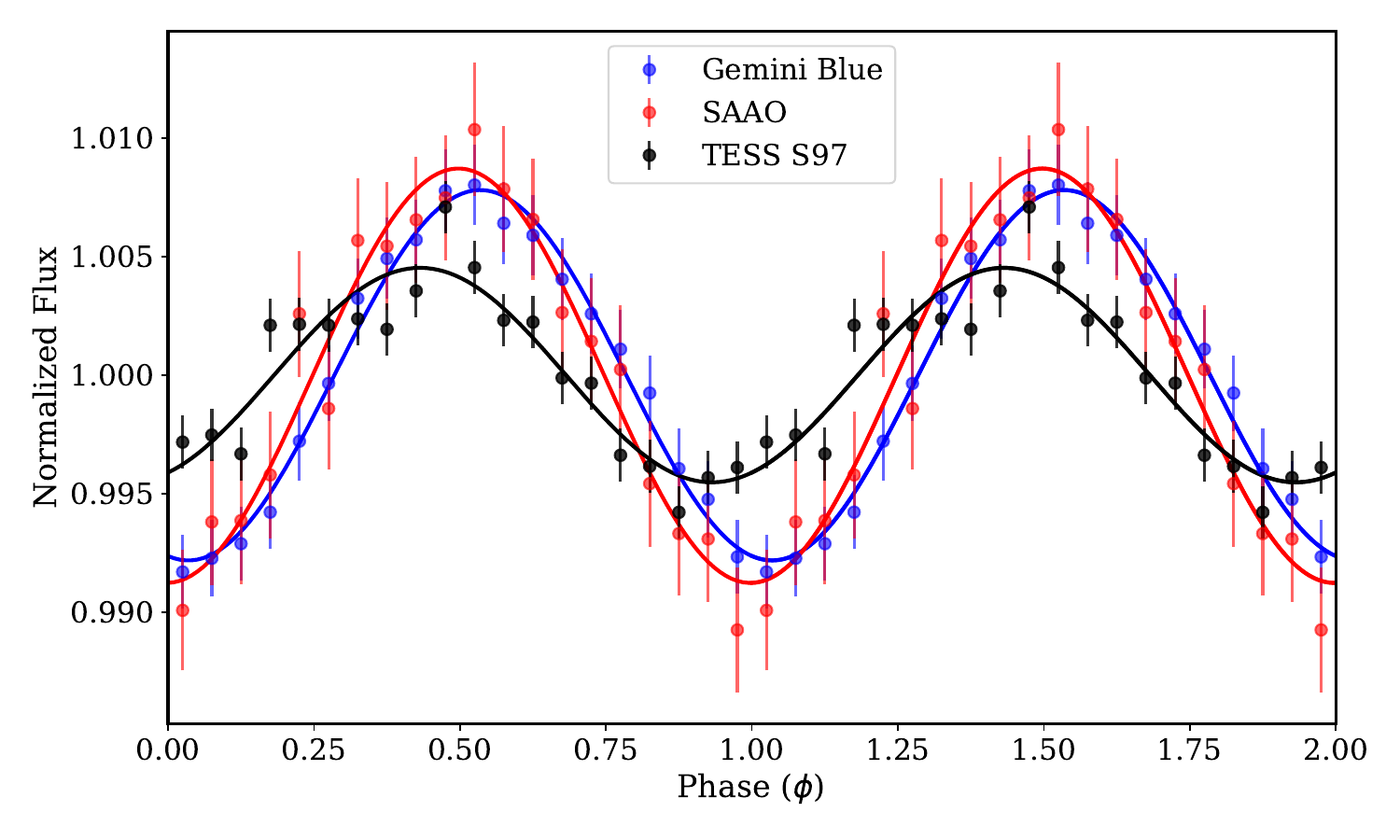}
    \caption{{\it Top:} Multi-instrument power spectrum analysis of \yzret. The plot compares the power spectra from GS, $SAAO$ and \tess. The \tess\ data are shown both within its nominal sampling regime (black) and its reflected power beyond the Nyquist (dashed green line) frequency (gray), illustrating the $P \approx 42.61$\,s alias. All instruments consistently resolve a dominant coherent peak at $f \approx 0.02652$ Hz ($P \approx 37.691$ s), definitively identifying the white dwarf spin period and resolving sampling ambiguities. {\it Bottom:} Phase-folded optical light curves of \yzret\ normalized to the mean flux. Two phase cycles are shown for clarity, with the modulation maximum manually aligned at $\phi = 0.5$. Solid lines represent the best-fit sinusoidal models for each dataset. The observed reduction in the \tess\ modulation amplitude relative to GS is fully consistent with the theoretical attenuation factor ($\approx 0.59$) derived from the finite integration time (sinc-function damping), due to the undersampling of the 20\,s cadence.}
      \label{fig:mirror}
\end{figure}

The LS periodograms of the $SAAO$ and GS data revealed a dominant signal at a frequency of $f_{\text{SAAO}} \approx f_{\text{GS}} \approx 0.02653$\,Hz (see Fig. \ref{fig:mirror}). While the discrepancy between this value and 
$f_{\text{TESS}}$ might initially suggest a rapid spin-down between November 2025 and March 2026 (Sect. \ref{sec:obs}), the frequencies are, in fact, related by a Nyquist reflection. Specifically, $f_{\text{TESS}}$ and $f_{\text{SAAO/GS}}$ satisfy the mirroring relationship:

\begin{equation}
f_{\text{TESS}} = f_{\text{Nyquist}} - (f_{\text{SAAO/GS}} - f_{\text{Nyquist}}) = 2f_{\text{Nyquist}} - f_{\text{SAAO/GS}}
\end{equation}

Given the 20-s cadence of the \tess\ observations which implies $f_{\text{Nyquist}} = 0.025$\,Hz), the fundamental frequency observed by $SAAO$ and GS is reflected across the Nyquist limit into the nominal sampling regime. As illustrated in Fig.~\ref{fig:mirror}, the signal at $0.02347$\,Hz in the \tess\ data is a sampling artifact (alias) of the true physical period at $37.691$\,s, which is correctly resolved by the higher-precision ground-based photometry.

Why does this happen?. The "Strobe Effect" can be illustrative. At every $TESS$ frame of 20\,s, the rotating body (in our interpretations, the white dwarf itself) has completed more than one full rotation per integration period (P$_{true}$=37.69131 $\pm$ 0.00001\,s). When we analyze the \tess\ light curve, we only see that phase wrap-around and thus the rotation seems slower ($P_{TESS}$=42.6105\,s). 

The nature of the 37.691\,s periodicity was investigated through a multi-instrument comparison of the pulse profiles (see Fig. \ref{fig:mirror}, bottom panel). We derived fractional semi-amplitudes ($A = [I_{\text{max}} - I_{\text{min}}] / 2 I_{\text{mean}}$) by performing weighted least-squares sinusoidal fits to the three phase-binned datasets. The high-cadence GS observations reveal an intrinsic semi-amplitude of $0.00778 \pm 0.00052$ in the $g'$-band. In the case of \tess\ Sector 97, the semi-amplitude is reduced to $0.00452 \pm 0.00035$, a factor approximately 0.58 times smaller, it is agreement with the theoretical damping factor of $0.59$ expected for a 37.691\,s signal sampled at a cadence, $\Delta t$ of 20\,s (sinc-function attenuation; ($A_{\tess} = A_{\text{true}} \times |\text{sinc}(\pi \Delta t / P)|$)). The stability of the modulation is further confirmed by the $SAAO$ data, which yield a semi-amplitude of $0.00873 \pm 0.00084$. This value is consistent with the GS semi-amplitude after accounting for the negligible sampling damping at a 5\,s cadence (factor $\approx 0.97$). 

\section{Discussion} \label{sec:disc}

In view of the evidence presented, we conclude that the optical emission of \yzret\ is characterized by a highly coherent modulation with a period of $37.69131 \pm 0.00001$ s. The stability of this periodicity over the five-month baseline between the \tess\ and GS-$SAAO$ observations suggests a fundamental origin, most likely associated with the rotation of a magnetic white dwarf. Such long-term frequency stability is inconsistent with typical Dwarf Nova Oscillations (DNOs), which are known for their transient nature and rapid decoherence on timescales of minutes to hours \citep{Warner2004}. Furthermore, the persistence of this signal in the current post-outburst high state \citep[VY~Scl phase, sometimes called "anti-dwarf novae"][]{Leach1999} is difficult to reconcile with the DNO phenomenon, further supporting the classification of \yzret\ as a fast-spinning Intermediate Polar (IP).

Alternative interpretations involving non-radial stellar pulsations, such as $g$-modes or $p$-modes, can also be excluded. Gravity-mode ($g$-mode) oscillations in pulsating white dwarfs (e.g., ZZ~Ceti stars) typically exhibit periods ranging from 100 to 1200\,s. Furthermore, $g$-modes are characterized by low-frequency stability, often showing significant amplitude modulations and mode-switching over timescales of days, which is inconsistent with the persistent and coherent pulse observed in \yzret. On the other hand, pressure-mode ($p$-mode) oscillations, which are driven by compressibility, are expected to have much shorter periodicities, generally below 10\,s for a typical white dwarf mass. The detected 37.691\,s period falls between these two regimes and seems unlikely to be related to those known intrinsic stellar pulsations mechanisms.

\section{Conclusions} \label{sec:concl}

We have discovered a fast optical periodicity in the post-nova \yzret\ and have definitively resolved the ambiguity surrounding its true value. By bridging the gap between \tess\ observations and high-speed ground-based photometry, we have identified a stable 37.691\,s white dwarf spin period. This discovery firmly classifies \yzret\ as an Intermediate Polar, making it one of the fastest-spinning magnetic white dwarfs known in a post-nova system. 

The stability of the spin frequency over 120 days, combined with the predictable amplitude suppression in the undersampled \tess\ data, provides a definitive identification with the white dwarf’s rotation. This resilient magnetic accretion regime continues to drive the system’s behavior even as it remains in a post-nova high state. The persistence of this spin signal -- despite the disruption of a classical nova eruption and the system’s subsequent return to a VY~Scl-type high state -- underscores the resilience of the magnetic accretion. 

The discovery of a 37.691\,s spin period in a system with an orbital period of 3.178\,hr \citep{Schaefer2022} places \yzret\ in a periods ratio, $P_{\text{spin}}/P_{\text{orb}} \approx 0.0033$, nearly two orders of magnitude smaller than that of typical IP ($P_{\text{spin}}/P_{\text{orb}} \sim 0.1$). Among a handful of novae already classified or proposed to be Intermediate Polars (see Table 1 by \citet{Orio2022} and references therein, and recent results in \citet{Orio2024}), \yzret\ is the one with the shortest period.

In a few other novae, several different short periods, even  as short as the proposed rotation period of \yzret\ (and intriguingly close to it, N~LMC~2009: 33\,s; KT~Eri and RS~Oph: $\simeq$35\,s) have often been detected in the supersoft X-ray source in outburst \citep[see][and numerous references therein]{Orio2022}. However, the shortest periods of only few tens of seconds have not been (or not yet) measured again in quiescence, so their root cause remains elusive and may be due to non-radial pulsations of some type.  

A further interesting implication is that, during the fireball phase or shortly thereafter, the rapidly expanding envelope may have reached or exceeded the break-up rotational velocity. At this stage, however, the wind was probably already active and extremely fast, driven not by common-envelope expansion or by the radiation-pressure mechanisms typically invoked in nova models \citep[see, among others,][]{Bath1976, Kato1989, Shen2022}, but instead by magnetic effects associated with rapid rotation \citep{Orio1992}.

In order for a magnetic or centrifugal rotator wind to occur, a primary mass loss trigger has to initiate the wind before it becomes sustained by the magnetic rotator. This initial trigger may be simply due to the thermonuclear runaway (TNR) shock wave in the envelope, especially if it is abundantly enriched with CNO nuclei by mixing. In a nova with a slowly rotating WD, the shock wave of the TNR may not last long enough to lose mass significantly  \citep[see discussion in][]{Starrfield2012, Starrfield2012b} but with such high rotation velocity and magnetic field as in IPs, the shock wave may be the trigger of the fast magnetic rotator wind. Alternatively, mass may start ``flying away'' in the expanding envelope because the radius is larger than the break-up radius.

We strongly encourage further studies at optical wavelengths to measure the system's orbital parameters, because {\it if we can determine that the inclination is NOT as low as it has been hypothesized, it is likely that the combined effect of rapid rotation and magnetic field in this nova caused the rapid, complete or almost complete expulsion of all the accreted envelope, quenching the burning and the related supersoft source unusually early.} This mechanism sustaining rapid mass loss has not been discussed again since it was proposed by \citet{Orio1992}, but it may indeed play a very important role in IP novae.

%%%%%%%%%%%%%%%%%%%%%%%%%%%%%%%%%%%%%%%%%%%%%%%%%%%%%%%%%%%%%%
\begin{acknowledgements}
%We acknowledge the anonymous referee for their careful reading and comments that significantly improved the manuscript. 
GJML is member of the CIC-CONICET (Argentina). AD was funded by the EU NextGenerationEU through the Recovery and Resilience Plan for Slovakia under the project No. 09I03-03-V04-00378. Some of the observations in the paper made use of the High-Resolution Imaging instrument Zorro. Zorro was funded by the NASA Exoplanet Exploration Program and built at the NASA Ames Research Center by Steve B. Howell, Nic Scott, Elliott P. Horch, and Emmett Quigley. Zorro was mounted on the Gemini South telescope of the international Gemini Observatory, a program of NSF NOIRLab, which is managed by the Association of Universities for Research in Astronomy (AURA) under a cooperative agreement with the U.S. National Science Foundation. on behalf of the Gemini partnership: the U.S. National Science Foundation (United States), National Research Council (Canada), Agencia Nacional de Investigación y Desarrollo (Chile), Ministério de Ciencia, Tecnología e Innovación (Argentina), Ministério da Ciência, Tecnologia, Inovações e Comunicações (Brazil), and Korea Astronomy and Space Science Institute (Republic of Korea). This paper uses observations made from the South African Astronomical Observatory (SAAO). This work is based on the research supported in part by the National Research Foundation of South Africa.
\end{acknowledgements}

% Please note that we have included the references below in
% order to compile the document, but we ask you to:

\bibliographystyle{aa}
\bibliography{listaref_MASTER}
%\begin{appendix}
%    To validate the choice of the AR model order $p$, we examined the Partial Autocorrelation Function (PACF) of the TESS light curves. The PACF typically showed a significant cutoff at lag $p$, which was objectively identified using the Akaike Information Criterion (AIC). As shown in Fig. [X], the resulting pre-whitening effectively flattened the power spectrum, removing the $1/f^\alpha$ red noise component characteristic of cataclysmic variable flickering
%\end{appendix}

\end{document}